\documentclass[aps,prb,reprint,showpacs,superscriptaddress,amsmath,amssymb]{revtex4-1}
\usepackage{graphicx}% Include figure files
\usepackage{dcolumn}% Align table columns on decimal point
\usepackage{bm}% bold math

\begin{document}

\title{Evolution of Co charge disproportionation with Na order in Na$_{x}$CoO$_{2}$}

\author{I.R.~Mukhamedshin}
\email{Irek.Mukhamedshin@kpfu.ru}
\affiliation{Institute of Physics, Kazan Federal University, 420008 Kazan, Russia}
\affiliation{Laboratoire de Physique des Solides, CNRS UMR 8502, Universit\'e Paris-Sud, 91405 Orsay, France, EU}
\author{A.V.~Dooglav}
\affiliation{Institute of Physics, Kazan Federal University, 420008 Kazan, Russia}
\affiliation{Laboratoire de Physique des Solides, CNRS UMR 8502, Universit\'e Paris-Sud, 91405 Orsay, France, EU}
\author{S.A.~Krivenko}
\affiliation{Institute of Physics, Kazan Federal University, 420008 Kazan, Russia}
\author{H.~Alloul}
\email{alloul@lps.u-psud.fr}
\affiliation{Laboratoire de Physique des Solides, CNRS UMR 8502, Universit\'e Paris-Sud, 91405 Orsay, France, EU}

\begin{abstract}
$^{59}$Co NMR experiments have been performed on single crystals of the
layered cobaltate Na$_{x}$CoO$_{2}$ with x=0.77 which is an antiferromagnet
with N\'eel temperature $T_{N}=22$~K. In this metallic phase six Co sites
are resolved in the NMR spectra, with distinct quadrupole frequencies $\nu
_{Q}$, magnetic shifts $K_{ZZ}$ and nuclear spin lattice relaxation rates $%
1/T_{1}$. Contrary to the $x=1/2$ or $x=2/3$ phases the 3D stacking of the
Na planes is not perfect for $x=0.77$ but this does not influence markedly
the electronic properties. We evidence that the magnetic and charge
properties of the Co sites are highly correlated with each other as $K_{ZZ}$
and $(1/T_{1})^{1/2}$ scale linearly with $\nu _{Q}$. The data analysis
allows us to separate the contribution $\nu_{Q}^{latt}$ of the ionic charges
to $\nu _{Q}$ from that $\nu _{Q}^{el}$ due to the hole orbitals on the Co
sites. We could extend coherently this analysis to all the known phases in
the Na cobaltate phase diagram. The variation with $x$ of $\nu _{Q}^{latt}$
is found to fit rather well numerical computations done in a point charge
model. The second term $\nu _{Q}^{el}$ allowed us to deduce the hole
concentration on the cobalts. These detailed experimental results should
stimulate theoretical calculations of the electronic structure involving
both the Co orbital configurations and DMFT approaches to take into account
the electronic correlations.
\end{abstract}


% 71.27.+a Strongly correlated electron systems; heavy fermions
% 76.60.-k Nuclear magnetic resonance and relaxation
% 61.66.-f Structure of specific crystalline solids
% 71.28.+d Narrow-band systems; intermediate-valence solids
% 75.20.Hr Local moment in compounds and alloys; Kondo effect, valence fluctuations, heavy fermions
% 76.60.Gv Quadrupole resonance

\maketitle

\section{Introduction}

Among the various families of correlated electron systems the layered
cobaltates Na$_{x}$CoO$_{2}$ are quite original due to the triangular
arrangement of the Co atoms in the CoO$_{2}$ layers. In the CoO$_{6}$
octahedra which build up the layered structure, the strong crystal field
induced on the Co lowers the $t_{2g}$ ionic levels with respect to the $%
e_{g} $ levels. This favors for the cobalt electronic structure the low spin
configurations in which the 5 to 6 electrons on the Co reside on the $t_{2g}$
ionic levels.\cite{SinghPRB61} This electronic structure is at the origin of
many singular physical properties, such as superconductivity, low $T$%
~Curie-Weiss (CW) behaviour, thermoelectric properties \textit{etc}.\cite%
{LangNFD}

As in many layered correlated electron systems, the variation of the
interlayer charge density, that is here the Na content, modifies the doping
of the CoO$_{2}$ layers, which permits to span a rich phase diagram.
Interestingly, initial investigations concluded that the stable Na contents
correspond to phases for which the Na atoms or equivalently the Na vacancies
are ordered.\cite{Zandbergen,Hinuma} This ordering of the dopants is at
variance with the case of cuprates for which the chemical dopants are quite
often disordered, and for which specific new phases are apparently favoured
by dopant order.

NMR/NQR experiments have proved so far to be excellent probes allowing to
evidence not only the Na atomic order but also that Co charge
disproportionation occurs. Indeed the $^{59}$Co NMR data easily give
evidence for the occurrence of nonmagnetic Co$^{3+}$ sites which correspond
to filled $t_{2g}$ levels. Such sites are absent for $x<0.5$ for which SC
occurs upon insertion of water. Their concentration progressively increases
with Na content for $x>0.5$ up to full Co$^{3+}$ content at $x=1$. But for $%
0.5<x<1$, the concentration of Co$^{3+}$ depends of the actual Na atomic
order.\cite{LangNFD}

Two specific ordered phases $x=0.5$ and $x=2/3$ could be studied in great
detail as their structure and NMR spectra could be resolved. For $x=0.5$ the
Na atoms are ordered in an orthorhombic superstructure commensurate with the
Co lattice.\cite{Na05structure} As a consequence a small charge
disproportionation into Co$^{3.5\pm \varepsilon }$ (with $\varepsilon <0.2)$
occurs on the two Co sites.\cite{Bobroff05} For the $x=2/3$ phase NMR/NQR
experiments have allowed the determination of both the 2D Na ordered
structure and the 3D stacking of Na/Co planes.\cite{EPL2009} The Co$^{3+}$
ions are stabilized on 25\% of the cobalt sites Co1 arranged in a triangular
sublattice. The holes are delocalized on the 75\% complementary cobalt sites
Co2 which display a planar cobalt kagom\'{e} structure, with an average
charge state close to Co$^{3.5+}$. The parameters of the Zeeman and
quadrupolar Hamiltonians determined for these Co2 sites have given evidence
that their electronic properties drive those of this $x=2/3$ phase.\cite%
{H67_CoNMR}

But so far it is not clear whether the original electronic properties are
driven solely by the quantum state in the 2D layer or by the Na order which
could help to pin this state. This question is raised as for $x\geq 2/3 $
four phases have the same CW magnetic behaviour above 100~K, but very
different ground states.\cite{EPL2008} For $x>0.75$ the detected 3D ordered
magnetic behaviour is suggestive of a role of the Na order.\cite%
{SugiyamaPRL2004,Mendels05}

Indeed all known phases with $x>0.62$ have ferromagnetic in plane properties
and display for $x>0.75$ a plane to plane antiferromagnetic (AF) order at
low $T$.\cite{BayrakciNa082,BoothroydNa075} To understand this difference in
the properties one needs to determine the details of the Na order and Co
charge disproportionation. This appears as a difficult task, as the control
of the synthesis of samples for large $x$ has been difficult so far as
contact with humid air leads to destruction of the phase purity and loss of
sodium content.\cite{JETP4phasesNQR} Progresses on that respect have been
done recently by selecting one of the most reproducible phase with $T_{N}=22$%
~K for which we could synthesise single phase crystals. From the $^{23}$Na
NMR study we could characterize the Na in plane order and could as well
evidence by $^{59} $Co NMR the existence of at least six Co sites, three of
them with quite distinct magnetic properties.\cite{Co77JETP}

This large number of sites opens a quite interesting situation permitting to
search for the correlation of the Co charge order with the magnetic
properties. For that we need to get accurate determinations of the
quadrupole data on each of the Co sites in order to try to distinguish the
contributions to the EFG of the static and mobile charges. The detailed
investigation and analysis of the $^{59}$Co NMR spectra is therefore the
main aim of the present paper, task which has been eased by the extensive
technical expertise we have developed in our previous study of the $x=2/3$
phase.\cite{H67_CoNMR}

After a brief recall in Sec.~\ref{Sec:AppSamples} of the experimental
techniques, in Sec.~\ref{Sec:Co77NMRSpectra} we perform detailed analysis of
the $^{59}$Co NMR spectra taken on single crystal samples in high magnetic
field, which permits us to identify the quadrupole frequencies $\nu _{Q}\ $%
for the six cobalt sites (we compare the $^{59}$Co NMR spectra taken on
powder and single crystal samples in Appendix~\ref{AppCoSpectra}). The
magnetic shift, $T_{1}$ and $T_{2}$ data are used to reveal the magnetic
properties detected by the given Co atomic sites. 

In Sec.~\ref{Sec:Correlations} we establish linear correlations
between the quadrupolar and magnetic properties, which is an evidence for
the variation of charge disproportionation on the Co sites, as discussed in
Sec.~\ref{Sec:Discussion}. There, the analysis of the data permits us to
separate the ionic contribution to the EFG with respect to the cobalt on
site delocalized charge contribution and therefore to evaluate the charge
disproportionation on the diverse sites of the structure.

The extension of this analysis to the entire phase diagram of Na
cobaltates permits us then to get a coherent picture of the evolution with
Na content of the various parameters considered above. This allows us then
to discuss the electronic structure questions which remain to be solved from
the theory standpoint to describe faithfully the experimental results
established so far.

\section{Samples and experimental techniques}

\label{Sec:AppSamples}

Experiments on sodium cobaltates Na$_{x}$CoO$_{2}$ with 22~K N\'{e}el
temperature have been performed on both oriented powders and single crystal
samples. The detailed description of the preparation methods and main
characteristics of the powder sample and single crystal SC1 have been given
in Ref.~\onlinecite{Na077prb}. Also a series of sodium cobaltates Na$_{x}$CoO%
$_{2}$ with $x\approx 0.8$ were grown independently by the floating zone
technique in Kazan Federal University. Using an electrochemical Na
de-intercalation method we reduced the sodium content in the as-grown
crystals down to that of the pure phase with 22~K N\'{e}el temperature and $%
x\approx 0.77$.\cite{Co77JETP} One of them was used in this study - below we
quote it as single crystal SC4.

The $^{59}$Co NMR measurements were done using home-built coherent pulsed
NMR spectrometers. NMR spectra were taken with $\pi /2-\tau -\pi /2$ or $\pi
/2-\tau -\pi $ radio frequency (rf) pulse sequences. Spectra obtained by
both sequences were quite similar, the only difference being the relative
intensities of the quadrupolar satellites and the central line. The usual $%
\pi /2$ pulse length was about 2-2.5~$\mu s$ depending on the coil size and
the minimum practical $\tau $ value used in our experiments was 5~$\mu s$.
The $^{59}$ Co NMR spectra were taken either in fixed field or in sweep
field mode. In fixed magnetic field $B_{0}$ the NMR spectra were recorded
point by point by varying the spectrometer frequency in equal frequency
steps. In sweep field mode, the NMR spectrometer frequency $\nu _{0}$ was
fixed and the spectra were taken in an external magnetic field which was
varied in equal steps. In both modes the full NMR spectra were then
constructed using a Fourier mapping algorithm.\cite{Clark,Bussandri}

\section{$^{59}$C\lowercase{o} NMR spectra}

\label{Sec:Co77NMRSpectra}

Generally in solid state NMR an atomic nucleus with spin $\vec{I}$,
quadrupole moment $Q$ and gyromagnetic ratio $\gamma $ has its spin energy
levels determined by the Zeeman interaction with the applied magnetic field $%
\vec{H}=B_{0}/\mu _{B}~$and the quadrupolar interaction with the electric
field gradient (EFG) on the nucleus site. The total Hamiltonian can be
written as:\cite{Abragam,Slichter} 
\begin{equation}
\mathcal{H}=\mathcal{H}_{Z}+\mathcal{H}_{Q}=-\gamma \hbar \vec{I}(1+\hat{K})%
\vec{H}+\frac{eQ}{2I(I-1)}\vec{I}\hat{V}\vec{I}.  \label{eq:Hamiltonian}
\end{equation}
In this Hamiltonian, the magnetic shift tensor $\hat{K}$ and the EFG tensor $%
\hat{V}$ reflect the physical properties of the studied compound and are
linked with the local structure. In the case where their principal axes $%
(X,Y,Z)$ coincide, the Zeeman and quadrupolar Hamiltonians can be
re-written: 
\begin{eqnarray}
&&\mathcal{H}_{Z}=-\gamma \hbar \sum_{\alpha }I_{\alpha }(1+K_{\alpha \alpha
})H_{\alpha };  \nonumber \\
&&\mathcal{H}_{Q}=\frac{eQ}{2I(2I-1)}\sum_{\alpha }V_{\alpha \alpha
}I_{\alpha }^{2}  \label{eq:Hamilt2} \\
&&\;\;\;\;=\frac{h\nu _{Q}}{6}\left[ 3I_{Z}^{2}-I(I-1)+\eta
(I_{X}^{2}-I_{Y}^{2})\right],  \nonumber
\end{eqnarray}%
where $\alpha =X,Y,Z$. Traditionally the principal axes of the EFG tensor
are selected as $\left\vert V_{ZZ}\right\vert \geq \left\vert
V_{YY}\right\vert \geq \left\vert V_{XX}\right\vert $. In that case the
quadrupolar frequency 
\begin{equation}
\nu _{Q}=\frac{3eQV_{ZZ}}{2I(2I-1)h}  \label{eq:NuQ}
\end{equation}
corresponds to the largest principal axis component $V_{ZZ}$ of the EFG
tensor and $\eta =(V_{XX}-V_{YY})/V_{ZZ}$ reflects the asymmetry of the EFG
tensor.

For a single Co site in the cell the simplest $^{59}$Co NMR spectrum is
detected when the principal axis of the EFG tensor $Z$ coincides with the
direction of the external applied field $H$. In first order perturbation
theory, it consists then of seven equally spaced lines: a central line which
corresponds to the $-\frac{1}{2}\leftrightarrow \frac{1}{2}$ transition and
6 quadrupolar satellites corresponding to the other $m\leftrightarrow (m-1)$
transitions which are symmetrically arranged around the central line. The
frequency splitting between lines in the spectrum is equal to the
quadrupolar frequency $\nu _{Q}$. The position of the central line in the
spectrum is determined by the $K_{ZZ}$ component of the magnetic shift
tensor.

\begin{figure}[tbp]
\includegraphics[width=1.0\linewidth]{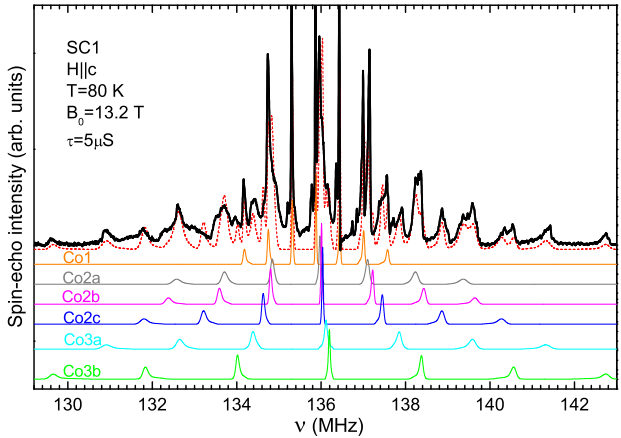}
\caption{(Color online) Experimental $^{59}$Co NMR spectrum (top thick black
line) of the SC1 sample measured in the fixed magnetic field $B_{0}\parallel
c$. The simulations of the spectra for 6 distinct Co sites NMR signals are
shown by colour lines. The parameters used in the simulations are presented
in Table~\protect\ref{tab:Co77Params}. The sum of the simulated cobalt
spectra with the weight coefficients from Table~\protect\ref{tab:Co77Params}
is shown by a red dotted line.}
\label{FigCo77HFSpec}
\end{figure}

In Fig.~\ref{FigCo77HFSpec} an example of $^{59}$Co NMR spectrum with good
resolution taken in high applied field $B_{0}\parallel c$ is shown for the
SC1 sample. One can see that there is a full set of distinct satellites
which can be sorted out knowing that each set of six lines are separated in
frequency by a specific $\nu _{Q}$ for each Co site

\subsection{Co1 sites NMR spectra}

As we have shown in our early studies,\cite{CoPaper} the $^{59}$Co NMR
spectra in cobaltates strongly depend on the time interval $\tau $ between
rf pulses, as the signals for the distinct sites which contribute to the NMR
spectrum have different nuclear relaxation parameters. We have used that at
length to separate the spectra for the sites with long and short transverse
relaxation times $T_{2}$. The existence of such a $T_{2}$ differentiation in
the $T_{N}=22$~K phase has already been demonstrated in Ref.~%
\onlinecite{Co77JETP} and can be seen in Appendix~\ref{AppCoSpectra} (Fig.~%
\ref{FigCo77Co40KHc} and Fig.~\ref{FigCo77Co40KHab}).

In the long $\tau $ spectrum the only detected signal contribution is that
for the sites with the longest $T_{2}$. We traditionally called the
corresponding cobalt sites as Co1 in most phases studied so far by NMR, and
associated them with Co sites which are nearly Co$^{3+}$, for which the six $%
t_{2g}$ sublevels are fully filled.

\begin{figure}[tbp]
\includegraphics[width=0.9\linewidth]{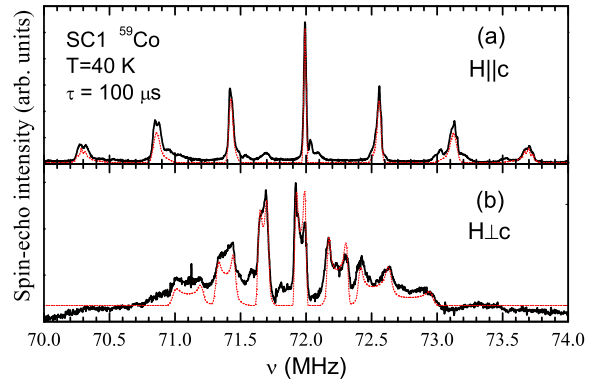}
\caption{(Color online) $^{59}$Co NMR spectra of SC1 sample measured at long
time interval $\protect\tau $ = 100$\protect\mu $s between rf pulses in the
two directions of the applied magnetic field. The simulation of the spectra
of Co1 are shown by the red dotted lines. See text for details.}
\label{FigCo77Co1Spectra}
\end{figure}

In Fig.~\ref{FigCo77Co1Spectra} we show examples of long-$\tau $ spectra of
the SC1 sample for $H\parallel c$ and $H\perp c$. The $H\parallel c$
spectrum displays clearly the expected seven lines spectrum which permits an
easy determination of $K_{ZZ}=2.36~\%$ and $\nu _{Q}=0.57$~MHz from the
computer fit shown in Fig.~\ref{FigCo77Co1Spectra}(a). For $H\perp c$, the
single crystal and aligned powder sample spectra are similar and some of
them are shown in Fig.~\ref{FigCo77Co40KHab} of the Appendix~\ref%
{AppCoSpectra}. Therefore all $XY$ orientations are seen in the single
crystal samples, which explains the complex lineshape of Fig.~\ref%
{FigCo77Co1Spectra}(b). The latter reflects the existence of an anisotropy
of the EFG characterized by the asymmetry parameter $\eta $. Computer
simulation assuming a $XY$ random disorder permits us to determine $\eta
\simeq 0.15$ from the best fit shown by a red dotted line in Fig.~\ref%
{FigCo77Co1Spectra}(b) (there a small broad background signal contribution
can be attributed to the residual signals of the short $T_{2}$ Co sites in
the structure). These simulations permitted us to evidence a small in-plane
asymmetry of the magnetic shift tensor components for Co1 sites: $%
K_{XX}=2.25~\%$ and $K_{YY}=2.36~\%$. We further found that the in-plane $XY$
principal axes of the magnetic shift and quadrupolar tensors are distinct
and rotated by about $25^{\circ}$ with respect to each other.

\subsection{Simulation of the spectra for the other Co lines}

We cannot do the same as for Co1 (Fig.~\ref{FigCo77Co1Spectra}) for the
other Co sites, as their spectra overlap and their $T_{2}$ are similar. This
is certainly impossible for $H\perp c$ (see Appendix~\ref{AppCoSpectra}),
but even for $H\parallel c$ we have to resort on a computer simulation to
fit the spectrum using the methods we have reported previously for $x=2/3$.%
\cite{H67_CoNMR} We could match the six quadrupole satellites associated
with each component of the central line for which $K_{ZZ}$ has been obtained.%
\cite{Co77JETP} In all studied samples at least six sets of Co quadrupolar
satellite lines could be distinguished by their $\nu _{Q}$ values. For each
Co site the largest splitting between satellites is observed when $%
H\parallel c$, which evidences that the principal axis $Z$ of the EFG tensor
is close to the crystallographic $c$ axis for all cobalts. Using the value
of $\nu _{Q}$ for each site we could compute its contribution to the
spectrum using the parameters listed in Table~\ref{tab:Co77Params}.

To simulate the broadening of the experimental quadrupole satellite lines we
combined the Lorentzian width $\Delta \nu $ used to fit the central line
spectrum in Ref.~\onlinecite{Co77JETP} with an additional Lorentzian
distribution of $\nu _{Q}$ with a full width at half maximum (FWHM) $\Delta
\nu _{Q}$ - see Table~\ref{tab:Co77Params}. The latter naturally reproduces
the increased linewidth of the external transitions with respect to the
internal ones. Finally the relative intensities of the six components have
been matched using those found for the central lines, which were confirmed
by accurate comparisons reported in Appendix~\ref{AppIntenAnalysis}. Those
allowed us to assign the number $I_{n}$ of sites associated with each Co
signal as given in Table~\ref{tab:Co77Params}

The corresponding fit summing the six contributions to the $^{59}$Co NMR
spectra is displayed as a dotted line in Fig.~\ref{FigCo77HFSpec}. Notice
that the discrepancies which occur are not fortuitous as we found them
highly reproducible in the spectra for different samples, as discussed in
Appendix~\ref{AppQuadrSatel}.

\begin{table}[btp]
\caption{Parameters used for computer simulations of the 6 contributions to
the $^{59}$Co NMR spectra shown in Fig.~\protect\ref{FigCo77HFSpec}. $K_{ZZ}$
and $\protect\nu _{Q}$ are the magnetic shift and the quadrupolar frequency
while $\Delta \protect\nu $ and $\Delta \protect\nu _{Q}$ are the widths of
their respective Lorentzian distributions. $I_{n}$ is the number of cobalts
corresponding to the given Co site in the simulated spectrum (total number
of cobalts is 13). The values of the longitudinal relaxation time $T_{1}$
and transverse nuclear magnetization relaxation time $T_{2}$ for each cobalt
site, measured at $T$=80~K are listed as well.}
\label{tab:Co77Params}%
\begin{ruledtabular}
\begin{tabular}{ccccccc}
                   & Co1   & Co2a  & Co2b  & Co2c  & Co3a & Co3b   \\
\hline
$K_{ZZ}$, \%       & 2.38  & 2.45  & 2.48  & 2.50  & 2.56  & 2.62  \\
$\Delta\nu$, kHz   & 16    &  44   &  20   & 21    & 81    & 41    \\
$\nu_Q$, MHz       & 0.568 & 1.14  & 1.217 & 1.42  & 1.745 & 2.19  \\
$\Delta\nu_Q$, kHz &  15   &  52   &  41   &  50   &   50  &  45   \\
$I_{n}$            &   3   &   4   &   1   &   2   &   2   &   1   \\
$T_1$, ms          &2.98(3)&0.91(5)&0.75(4)&0.71(4)&0.30(1)&0.17(1)\\
$T_2$, $\mu $s     & 147(4)& 45(2) & 47(2) & 44(3) & 19(1) & 16(1) \\
\end{tabular}
\end{ruledtabular}
\end{table}

\section{Correlation between charge and magnetic properties}

\label{Sec:Correlations}

The unfilled local charge, if it is localized as in a Mott insulator, is
directly linked with the ionic magnetism, while in a metal the unfilled
shell participates in the conductivity and magnetic properties of the
metallic state. Obviously, in our metallic systems the large variation of $%
\nu _{Q}$ for the various Co sites is related to the local Co orbitals that
participate in the metallic bands, as we have already seen for other $x$
values.\cite{H67_CoNMR} We have now in Table~\ref{tab:Co77Params} the $%
K_{ZZ} $ and $\nu _{Q}$ values which are respectively associated with
magnetic and charge properties for the six sites of the Co plane, so we can
test how they correlate to each other.

\subsection{Correlation between $K_{ZZ}$ and $\protect\nu_{Q}$}

\begin{figure}[tbp]
\includegraphics[width=0.9\linewidth]{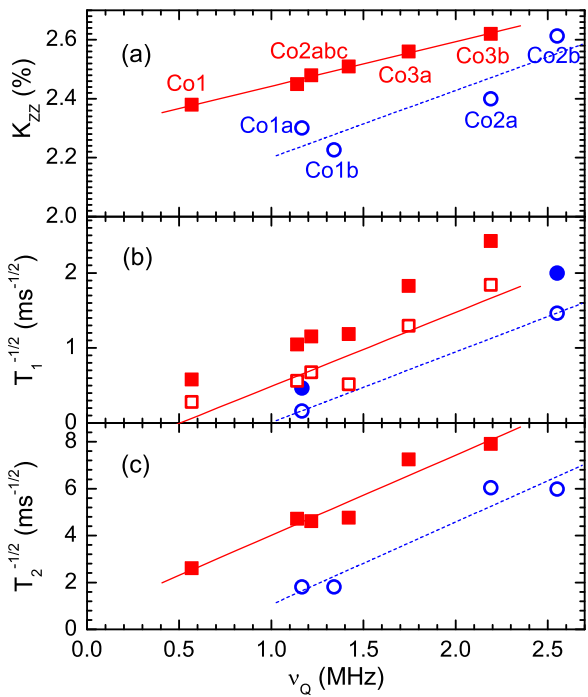}
\caption{(Color online) (a) $K_{ZZ}$, (b) $(1/T_{1})^{1/2}$ (closed symbols)
and $(1/T_{1spin})^{1/2}$ (open symbols) (c) $(1/T_{2})^{1/2}$ vs. $\protect%
\nu _{Q}$ for cobalt sites in $x=0.77$ phase (red squares) and $x=2/3$ phase
(blue circles). Linear fits are shown as eyeguides.}
\label{FigCo77Corr}
\end{figure}

Fig.~\ref{FigCo77Corr}(a), where $K_{ZZ}$ versus $\nu _{Q}$ values listed in
Table~\ref{tab:Co77Params} are plotted, evidence a simple linear correlation
between these two quantities. On the same figure we reported the data for
these quantities obtained for the four Co sites detected for $x=2/3$.\cite%
{H67_CoNMR} Though they do not scale as nicely with each other as for the $%
x=0.77$ case, the trend is quite similar and the Co1a and Co1b sites
correspond there to smaller $K_{ZZ}$ and $\nu _{Q}$ than for Co2a and Co2b
sites.

Conversely we have evidenced for the $x=2/3$ phase that $%
K_{XX},K_{YY}>K_{ZZ} $ on the Co2 magnetic sites.\cite{H67_CoNMR} It would
be important to determine if this happens as well for the magnetic sites of
the $x=0.77$ phase. For $x=2/3$ this was seen from direct shift measurements
in the $H\perp c$ direction. There we developed a full analysis which
permitted us to distinguish the $K_{XX},K_{YY}$ contributions for the two
fast relaxing sites of the structure Co2a and Co2b. As in the $x=2/3$ phase,
we can see a strong anisotropy of the shift in the $H\perp c$ spectra (see
Appendix~\ref{AppCoSpectra}), but those are too complicated to attempt such
a differentiation for the five fast relaxing Co sites for $x=0.77$.

\subsection{Correlation between T$_{1}$, T$_{2}$ and $\protect\nu_{Q}$}

We have demonstrated for $x=2/3$ that $T_{1}$ and $T_{2}$ data also allow us
to detect the anisotropy of magnetic properties of the Co2 sites.\cite%
{H67_CoNMR} We shall therefore use these quantities to further test the
correlation between the magnetic properties and $\nu _{Q}$ for the $x=0.77$
phase.

Due to the close values of the magnetic shift for all cobalts sites in the $%
H\parallel c$ direction (see Table~\ref{tab:Co77Params}), to avoid
cross-relaxation between sites we measured $T_{1}$ on selected quadrupolar
satellite transitions of different sites which are quite separated in
frequency - see Fig.~\ref{FigCo77HFSpec} as example. The details of the
measurements are given in Appendix~\ref{AppT1details} and the $T_{1}$ values
measured for the different sites are listed in Table \ref{tab:Co77Params}.
We do see there that $1/T_{1}$ increases markedly with $\nu _{Q}$.

Let us recall that, in systems with unpaired spins, the dominant $T_{1}$
process is due to local field fluctuations induced by the dynamics of the
local electronic magnetization.\cite{H67_CoNMR} For $H\parallel c$, this $%
T_{1}^{-1}$ is governed by the fluctuations of the transverse components of
the effective magnetic field induced at the nucleus. Thus it is driven then
by the $K_{XX}$ and $K_{YY}$ components of the magnetic shift tensor so that 
$T_{1Z}^{-1}\propto (K_{YY}^{2}+K_{XX}^{2})$. So, to probe the correlation
of the transverse spin shifts with the on site charge, it appears practical
to plot $(1/T_{1Z})^{1/2}$ versus $\nu _{Q},$ as done in Fig.~\ref%
{FigCo77Corr}(b). There one can see that a clear linear relation holds
amazingly for the six sites and can by no way be considered as fortuitous.
This plot means that the transverse spin shifts $K_{XX}$ and $K_{YY}$ do
scale as well with $\nu _{Q}$ or $K_{ZZ}$. Therefore the large $T_{1}^{-1}$
values found for the Co3 sites with large $\nu _{Q}$ values clearly confirm
the magnetic character of these cobalt sites.

The transverse relaxation rate $(1/T_{2})^{1/2}$ displays a similar scaling
as shown in the plot of Fig.~\ref{FigCo77Corr}(c), which means that it is
dominated by the $T_{1}$ process.

Although the scaling is clear, the ratio of $(1/T_{1})^{1/2}$ for Co3a and
Co3b is only about 1.3 and the three Co2 sites have similar $T_{1}$ values.
Therefore the rough distinction done from the $T_{2}$ data in Ref.~%
\onlinecite{Co77JETP} between the three types of sites Co1, Co2 and Co3 is
maintained. Co1 is close to nonmagnetic Co$^{3+}$, Co3a and Co3b sites are
the most magnetic cobalt sites in this x=0.77\ phase, and the remaining
three Co2 sites display an intermediate behaviour.

\section{Discussion}

\label{Sec:Discussion}

To explain a linear variation of the Knight shift $K_{ZZ}$ with $\nu _{Q}$
(Fig.~\ref{FigCo77Corr}(a) let us consider the significance of these
measured physical quantities.

The quadrupole frequency $\nu _{Q}$ is proportional to the largest component
of the EFG tensor $\hat{V}$ (Eq.~(\ref{eq:NuQ})). The components of $\nu _{Q}
$ and $\hat{V}$ can be written as the sums of two contributions 
\begin{eqnarray}
&&\nu _{Q}=\nu _{Q}^{latt}+\nu _{Q}^{el} \\
&&\hat{V}=(1-\gamma _{\infty })\ \hat{V}^{latt}+(1-R_{el})\ \hat{V}^{el}.
\label{eq:V}
\end{eqnarray}%
The first term in Eq.~(\ref{eq:V}), which gives the lattice contribution to $%
\nu _{Q}$ arises from all ion charges outside the ion under consideration,
enhanced by the core electrons of the atom by Sternheimer antishielding
factor $\gamma _{\infty }$. The second term in Eq.~(\ref{eq:V}) gives the
local electronic contribution to $\nu _{Q}$ and arises from unfilled
electron shells of the orbitals of the considered site. Those induce
distortions of the inner electronic orbitals embedded in the value of $%
(1-R_{el})$ which usually just slightly smaller than unity.

For an isolated ionic site in an insulator the EFG tensor for one hole on $%
t_{2g}$ orbitals would be expressed as\cite{AbragamBleaney,PRLKiyama} 
\begin{equation}
\hat{V}^{el}=-\frac{2}{21}\langle r^{-3}\rangle e\hat{q},  \label{eq:Vel}
\end{equation}%
where $\langle r^{-3}\rangle $ is the expectation value of $r^{-3}$ for the $%
3d$ electron and $e\hat{q}$ is the electric quadrupole moment of the $t_{2g}$
hole. The on-site spatial distribution of the hole is represented by the
second-rank tensor $\hat{q}$, given by the expectation value of the
operators $q_{\alpha \beta }\equiv \frac{3}{2}(L_{\alpha }L_{\beta
}+L_{\beta }L_{\alpha })-\delta _{\alpha \beta }L^{2}$ ($\alpha ,\beta
=x,y,z $) where $L$ is the orbital momentum.\cite{PRLKiyama}

In the metallic state constructed from the low spin ionic configuration of
the Co one has to take into account solely the contribution of the Co
orbitals involved in the metallic band up to the Fermi level, so that the on
site charge $\delta $, corresponding to the disproportionated charge in
chemistry language, has to be introduced as a multiplicative factor in Eq.~(%
\ref{eq:Vel}). For simplicity $\delta $ might be embedded in the definition
of $\hat{q}$, which would represent then the local charge concentration
weighted by an anisotropy of the angular distribution of electronic density.
This local charge term is important as it represents a direct on site
signature of the electronic structure. For instance for the Co$^{3+}$ low
spin state with a fully filled $t_{2g}$ multiplet the local electronic
structure is isotropic and should not contribute to $\hat{V}^{el}$ and $\nu
_{Q}^{el}$.

The magnetic shift tensor $\hat{K}$ reflects the additional magnetic fields
on the nuclear site which could be written in terms of the spin $\vec{S}$
and orbital $\vec{L}$ operators:\cite{AbragamBleaney,PRLKiyama} 
\begin{equation}
\hat{K}\vec{H_{0}}=-2\mu _{\mathrm{B}}\langle r^{-3}\rangle (\ \vec{L}%
-\kappa \vec{S}+\frac{2}{21}\hat{q}\vec{S}).  \label{eq:KH0}
\end{equation}%
Here $\mu _{\mathrm{B}}$ is the Bohr magneton and the orbital contribution
to the shift has been separated from the spin contribution.

The first term in Eq.~(\ref{eq:KH0}) describes the magnetic interaction
between the nuclear spin and the $d$-orbital moment. The second contribution
corresponds to the isotropic contact Fermi coupling of the nucleus spin to
the spin polarization of the inner $s$-electrons shells induced by the outer 
$d$ electrons (core polarization). The third term in Eq.~(\ref{eq:KH0})
describes the magnetic dipolar interaction between the nuclear spin and the
quadrupolar part $\hat{q}\vec{S}$ of the distribution of the $t_{2g}$-hole
spin density.\cite{AbragamBleaney}

It is important that both measured values $\nu _{Q}$ and $\hat{K}$ contain
terms proportional to the quadrupole moment of the $t_{2g}$ hole density
distribution $\hat{q}$ which involves as a coefficient the hole
concentration $\delta $ on the Co site. Therefore the linear variation with $%
\nu _{Q}$ found for the Knight shift $K_{ZZ}$ and $(T_{1})^{-1/2}$ just
reflects the fact that the hole content on the Co orbitals varies from site
to site in this $x=0.77$ phase. Also this observation means that the
hyperfine coupling, or the local magnetic $\chi $ scales with the on site
delocalized charge.

\subsection{Estimation of the charge disproportionation for the different Co
sites}

The second term in Eq.~(\ref{eq:V}) is clearly linked with the Co on site
hole occupancy, and does vanish in the Co$^{3+}$ low spin state with six
electrons filling the $t_{2g}$ orbitals. Such Co atoms being nonmagnetic, we
do expect their $T_{1}$ to become extremely long as observed experimentally.%
\cite{CoPaper,LangNa1,MHJulienNa1}

Therefore in the plot of Fig.~\ref{FigCo77Corr}(b) the pure Co$^{3+}$ state
should correspond to $(1/T_{1})^{1/2}$ $\rightarrow 0$ so that the
corresponding $\nu _{Q}=\nu _{0}\simeq 0.2~MHz$ should be solely due to the
lattice contribution $\nu _{Q}^{latt}$ to the EFG. We can therefore
anticipate that the quantity $\nu _{Q}-\nu _{0}$ is directly related on each
site with the local charge term in Eq.~(\ref{eq:V}).

It therefore measures the charge disproportionation on the different sites
of the structure, and we could write $\delta _{n}=A(\nu _{Q,n}-\nu _{0})$
for a site $n$ with quadrupolar frequency $\nu _{Q,n}$ and charge $3+\delta
_{n}$. Knowing the number $I_{n}$ of sites $n$ in the unit cell the charge
neutrality implies 
\begin{equation}
A\sum I_{n}(\nu _{Q,n}-\nu _{0})=(1-x)\sum I_{n}.  \label{eq:A}
\end{equation}%
The data for $\nu _{Q,n}$ and $I_{n}$ obtained from the signal intensity
analysis given in Table~\ref{tab:Co77Params} therefore permit us to
determine $\delta _{n}$ values listed in Table~\ref{tab:HoleConcentration}.
These estimates would indicate that the disproportionation ranges from Co$%
^{3.08+}$ for the Co1 sites, with an average of Co$^{3.24+}$ on the Co2
sites and an average of Co$^{3.40+}$ for the Co3 sites.

We did for $x=2/3$ an equivalent analysis based on the values of $\nu _{Q}$
for the Co sites of the kagom\'{e} structure.\cite{H67_CoNMR} It is
reproduced in Fig.~\ref{FigCo77Corr}(b), which gives $\nu _{0}\sim 0.7$~MHz
and the $\delta _{n}$ values listed in Table~\ref{tab:HoleConcentration}.

In fact, for completeness we have to take into account that the measured $%
T_{1}$ involves an orbital contribution $1/T_{1orb}$ which we evidenced
explicitly for $x=2/3$ in Ref.~\onlinecite{H67_CoNMR}. As the $\delta _{n}$
values (Table~\ref{tab:HoleConcentration}) varies in the same range for both 
$x=0.77$ and $x=2/3$ phases and $1/T_{1}$ values for cobalts are also quite
close, we assumed that for the $x=0.77$ phase the $1/T_{1orb}$ versus $\nu
_{Q}$ dependence has the same slope as for the $x=2/3$ phase which allowed
us to estimate $1/T_{1orb}$ for all Co sites in the $x=0.77$ phase. In Fig.~%
\ref{FigCo77Corr}(b) we report the spin contributions $1/T_{1spin}$ to the
relaxation for both phases after subtracting the orbital term. We find that
using $(1/T_{1spin})^{1/2}\rightarrow 0$ only induces a $\approx$~0.3~MHz
increase of $\nu _{0}$ values for both 0.77 and 2/3 phases. As can be seen
in Table~\ref{tab:HoleConcentration} this does not change significantly the $%
\delta_{n}$ values except for the Co1 sites which become closer to Co$^{3+}$.

The slope of $(1/T_{1spin})^{1/2}/\nu _{Q}$ is found quite similar in Fig.~%
\ref{FigCo77Corr}(b) for both samples. This can be ascribed to the analogous 
$T$ variations of $1/T_{1}$ found for such phases above 80~K.\cite{EPL2008}
Finally a similar analysis of low $T$ NQR data taken on the O71 sample,
which is detailed in Appendix~\ref{AppO71} also gives a distinction between
two types of sites and very similar values as for the two sites for $x=2/3$.
So the comparison of these three cases leads us to the conclusion that for $%
x<0.75$ the disproportionation to nearly nonmagnetic Co$^{3+}$ and magnetic
Co$^{\approx 3.5+}$ sites happens. The disproportionation appears more
complex for the $x=0.77$ phase as apart Co$^{3+}$ two differently charged
states are involved.

\begin{table}[tbp]
\caption{Hole concentration $\protect\delta $ deduced from the analysis of
the correlation between $(1/T_{1})^{1/2}$ and $\protect\nu _{Q}$ data for
different cobalt sites with charge states Co$^{3+\protect\delta }$ in $x=2/3$
and $x=0.77$ phases. In the tables, the upper values of $\protect\nu _{0}$
are obtained from the raw data and the lower ones after subtracting the
orbital contribution to $1/T_{1}$, as displayed in Fig.~\protect\ref%
{FigCo77Corr}(b)}
\label{tab:HoleConcentration}%
\begin{tabular}{c|c|c|c|c}
\multicolumn{5}{c}{$x$=2/3} \\ 
$\nu_0$, MHz & Co1a & Co1b & Co2a & Co2b \\ \hline
0.7 & 0.11 & 0.15 & 0.35 & 0.43 \\ 
1.0 & 0.05 & 0.10 & 0.35 & 0.46%
\end{tabular}
\newline
\begin{tabular}{c|c|c|c|c|c|c}
\multicolumn{7}{c}{$x$=0.77} \\ 
$\nu_0$, MHz & Co1 & Co2a & Co2b & Co2c & Co3a & Co3b \\ \hline
0.2 & 0.08 & 0.21 & 0.23 & 0.27 & 0.35 & 0.45 \\ 
0.5 & 0.02 & 0.20 & 0.23 & 0.29 & 0.39 & 0.53%
\end{tabular}%
\end{table}

\subsection{Evolution of the ionic contribution to $\protect\nu _{Q}$ in the
phase diagram}

One can expect that the $\nu _{Q}^{latt}$ contribution to $\nu _{Q}$ changes
with Na content and order. In rectangular coordinates the components of
corresponding $\hat{V}$ $^{latt}$ tensor can generally be calculated in a
point charge model 
\begin{equation}
V_{\alpha \beta }^{\mathrm{latt}}=\sum_{i}q_{i}\left( \frac{3\alpha
_{i}\beta _{i}-r_{i}^{2}\delta _{\alpha \beta }}{r_{i}^{5}}\right) ,
\end{equation}%
where $q_{i}$ is the charge of the ion $i$, located with respect to the
probe nucleus at the vector $\vec{r}_{i}$ with components $\alpha _{i}$ and $%
\beta _{i}$ with $\alpha ,\beta =X,Y,Z$.

So $\nu _{Q}^{latt}$ can be estimated in a first step without taking into
account the Na order and the Co charge disproportionation. We performed a
simple point charge calculation, assuming a homogeneous distribution of
ionic charge on the Na sites (that is a charge $x$ on every Na2 site),
charges -2 on all O sites and $4-x$ on each Co. \footnote{%
In the point charge model calculations for the $^{59}$Co quadrupole moment
Q=0.42 and Sternheimer antishielding factor $\gamma =-7$ were used.} This
resumes into calculating the contributions $\nu _{QA}$ (Na2:0; O:-2; Co:4)
which represents the contribution of the non-charged CoO$_{2}$ layers and $%
\nu _{QB}$ (Na2:$x$; O:0; Co:-$x$) representing the Na and hole
contributions to the EFG.

\begin{figure}[tbp]
\includegraphics[width=1.0\linewidth]{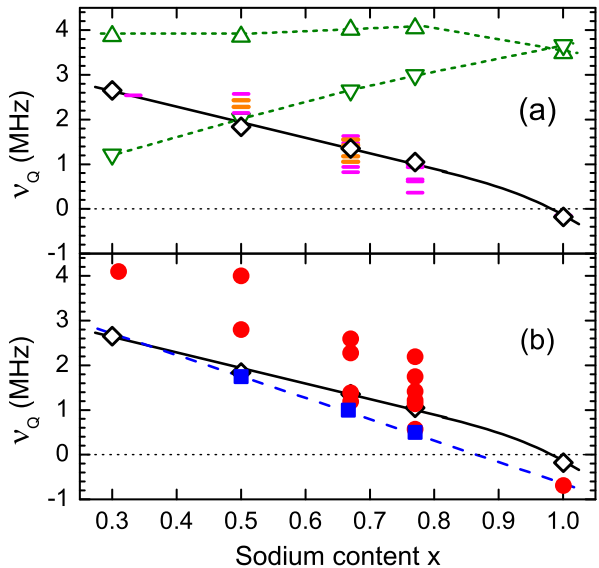}
\caption{(Color on line) (a) The lattice contribution to the quadrupole
frequency variation $\protect\nu _{Q0}^{latt}$ (black open diamonds and
black line) computed for a uniform distribution of Na and Co sites. It
results from the contributions $\protect\nu _{QA}$ and $\protect\nu _{QB}$
(green open up and down triangles correspondingly). Using the actual Na
order yields the differentiation of $\protect\nu _{Q}^{latt}$ values on the
Co sites given by horizontal magenta ticks. Including the Co charge
disproportionation slightly reduces this differentiation (orange ticks) and
only slightly modifies the average magnitude of the computed $\protect\nu %
_{Q}^{latt}$ values with respect to the uniform case. (b) Comparison of the
experimental $\protect\nu _{Q}$ data (red full circles) with point charge
calculations. The $\protect\nu _{0}(x)$ values deduced from the analyses of
Fig.~\protect\ref{FigCo77Corr}(b) (blue closed squares and dashed line)
parallels the variation of the computed data. See text for details.}
\label{FigNEW}
\end{figure}

The calculated $\nu _{Q0}^{latt}=\nu _{QA}+\nu _{QB}$ on the cobalt sites is
plotted versus $x$ in Fig.~\ref{FigNEW} using the cell parameters known from
refinements of x-ray data.\cite%
{Na03structure,Na05structure,H67NQRprb,Huang09,LangNa1}

To further test the validity of our analyses of the data we need to
appreciate the soundness of the evolution with Na content of $\nu _{0}$
which we have also plotted in Fig.~\ref{FigNEW}.

There one can see that the variation with $x$ of $\nu _{Q0}^{latt}$ has a
similar regular decrease with $x$ as that displayed by $\nu _{0}(x)$, though
the computed values are slightly larger. Of course the existence of Na order
and charge disproportionation might justify this difference. As the Na order
unit cell is perfectly known for $x=1/2$ and $x=2/3$, we therefore could
compute $\nu _{Q}^{latt}$ for those phases. The computed values on the
distinct Co sites do not correspond to a single value and deviate from $\nu
_{Q0}^{latt}$. However the differentiation between the computed $\nu
_{Q}^{latt}$ values is reduced if one takes also into account the Co charge
disproportionation determined above from the analysis of the experimental $%
\nu _{Q}$ values.

We computed also the expected $\nu _{Q}^{latt}$ for a hypothetical
triangular arrangement of Na ions on Na2 sites for $x=1/3$. Similarly for
the $x=0.77$ phases we calculated $\nu _{Q}^{latt}$ for different stacking
along $c$-axis of 2D Na order pattern established in Ref.~\cite{Na077prb}
Despite the uncertainty in the 3D stacking the trend in the variation of $%
\nu _{Q}^{latt}(x)$ remains similar.

In Fig.~\ref{FigNEW} we have reported as well the experimental $\nu _{Q}$
values for the various Co sites identified for the studied Na contents. As
expected these data are mostly above the computed values, whatever the
approximation done, as the local charge contribution $\nu _{Q}^{el}$ of the
unfilled Co orbitals does of course contribute to $\nu _{Q}$.

One puzzling result is that for $x=1$, the calculated value of $\nu
_{Q}^{latt}$ is also found below the experimental result, while in this
insulating band case one does not expect any $\nu _{Q}^{el}$ contribution.
When trying to clarify that point we noticed that in the homogeneous charge
model, $\nu _{QA}$ and $\nu _{QB}$ are of opposite sign and nearly
compensate each other for $x=1$. This explains the difficulty to fit the $%
x=1 $ data, as the Co-O hybridization might yield deviations from point
charge calculations for the contribution $\nu _{QA}$ of the CoO$_{2}$ slabs.

We may notice that for most Na contents the computed values for $\nu
_{Q}^{latt}$ are always slightly above the actual $\nu _{0}(x)$ values, so
that one might consider that $\nu _{QA}$ is always slightly overestimated.
We are then led to conclude that $\nu _{0}(x)$ values are rather good
determinations of $\nu _{Q}^{latt}$. This could mean that the actual value
of $\nu _{Q}$ for $x=1$ would be negative, as the sign of the EFG is not
accessible experimentally. In such a case the variation of $\nu _{0}(x)$
would smoothly extend linearly up to $x=1$ as shown in Fig.~\ref{FigNEW}.

All this gives some weight to the hypothesis of Eq.~(\ref{eq:A}) that a
single $\nu _{0}$ can be used for all Co sites to describe the
disproportionation for a given Na content. This approximation is certainly
not fully valid for the most stable ordered Na phases, such as $x=2/3$ for
which the differentiation between Co sites is slightly influenced by the Na
order.

\subsection{Hole contribution to $\protect\nu _{Q}$ on the Co nuclei}

The variation of $\nu _{Q}$ with the local charge $\delta $ is given by $%
d\nu _{Q}/d\delta =A^{-1}$ (A is defined in Eq.~(\ref{eq:A})). For the $\nu
_{0}$ values corresponding to $(1/T_{1})^{1/2} \rightarrow 0$ we obtain
4.3~MHz and 4.5~MHz per on site charge for $x=2/3$ and $x=0.77$
respectively, while we do obtain 3.4~MHz and 3.2~MHz for the $\nu _{0}$
values taken for $(1/T_{1spin})^{1/2} \rightarrow 0$. Recognizing the
roughness of our model one could retain a reasonable value of 3.3~MHz per
hole on Co site.

For phases with the $1/2<x<2/3$ the weak $T$ variations of the NMR shifts
and the lower resolution of the $^{59}$Co powder spectra did not permit to
get accurate data in which all sites could be identified.\cite{LangNFD} But
it has been shown that the concentration of non magnetic Co1 like sites
decreases with decreasing $x$ and vanishes at $x=0.5$. In that well ordered
phase the two sites differentiated correspond to Co$^{3.5+\varepsilon }$ and
Co$^{3.5-\varepsilon}$.\cite{Bobroff05} Providing we keep the same $d\nu
_{Q}/d\delta $=3.3~MHz per hole as for higher $x$, the 2.8~MHz and 4~MHz
measured $\nu _{Q}$ values allow us to estimate $\varepsilon =0.18$. This
correspond then to $\nu _{0}(0.5)=1.75$~MHz which is quite close to that
computed for $\nu _{Q}^{latt}(x=1/2)$ - see Fig.~\ref{FigNEW}. Therefore, we
find that the computed ionic contribution $\nu _{Q}^{latt}(x)$ is reliable,
and may be slightly improved by using the value $\nu _{0}(x)$ deduced from
the analysis of the charge disproportionation.

As for the extra contribution due to the local charge $\nu _{Q}^{el}$, it
corresponds to about 3.3~MHz per on site charge for the variations of $\nu
_{Q}$ on the various sites of the $x=2/3$ and $x=0.77$ phases, but also is
compatible with the data on the $x=1/2$ phase. The occurrence of such local 
charge contribution implies that the local charge is not evenly distributed 
on the Co $t_{2g}$ orbitals.

Let us now evaluate the EFG which should results from a single electron or
hole residing in one $t_{2g}$ orbital, employing Eqs.~(\ref{eq:NuQ}) and (\ref%
{eq:V}). The magnitude of the largest component of the quadrupole moment
tensor for the orbital state was calculated as $\left\vert \hat{q}%
\right\vert \approx 6$.\cite{PRLKiyama} In the free-ion limit for Co$^{4+}$
the coefficient $\langle r^{-3}\rangle$ has been estimated as $7.421~\text{%
a.u.}$\cite{AbragamBleaney} Because of a hybridization with the ligands in
the sodium cobaltates this value becomes smaller, and $\langle r^{-3}
\rangle=5~\text{a.u.}$ could be used as a realistic approximation.
Substituting these quantities into Eqs.~(\ref{eq:NuQ}) and (\ref{eq:V}) and
taking $R_{el}\ll 1$, we obtain $\nu _{Q}^{el}\approx $ 20~MHz per single
orbital for $^{59}$Co. Similar values of EFG could remain in the case of 
a robust orbital order in the $d$-lattice. Such ordering could appear when 
the orbitals are locked by a local crystal field with a broken cubic 
symmetry.\cite{KHA2005} In the cobaltates such field could originate from a 
deformation of the Co-O octahedra and/or from a low-symmetric sodium 
structure. For example a trigonal distortion of the oxygen octahedra induces 
a $a_{1g}$ state of the Co$^{4+}$ sites.\cite{KHA2008} 

The calculated quadrupolar frequency is about six times larger then the 
corresponding experimental value $3.3~\text{MHz}$ per Co site established in 
our analysis above. This suggest that in the cobaltates the holes are 
distributed over the three orbital states within the $t_{2g}$-shells due to 
the thermal/quantum fluctuations.\cite{KHA2008} Such effect 
could explain the recovery of the isotropy of the local wave 
functions.\cite{KHA2005,KRI2012} A plausible possibility for the reduction 
could come from the band structure of the holes in the triangular CoO$_{2}$ 
layers, which would intermix the three planar $t_{2g}$ orbitals 
($xy, yz, xz$).\cite{KoshibaeMaekawa,Lechermann2011} The quantitative description of this 
situation is a challenge for future theoretical developments.

\section{Conclusion}

We have evidenced here that the $^{59}$Co NMR spectrum reflects
in the paramagnetic state the diverse Co sites pertaining to the Na ordered
phases in the Na cobaltates. We demonstrated that, for the $x=0.77$ 
phase with $T_{N}=22$~K, the Co charge disproportionation
differentiates magnetically three type of sites, while for $2/3<x<0.75$ 
only two types of magnetic behaviour occur. Furthermore we could
establish here that for $x=0.77$ the charge disproportionation and
the local magnetic behaviour on the Co sites are very well correlated.

The atomic structure of this $x=0.77$ phase corresponds to a Na order which involves Na tri-vacancies and triangles of Na1 sites rather
than the di-vacancies and isolated Na1 sites found for lower $x$.\cite{Na077prb} We did also find here that the stacking of the Na order is far from being perfectly locked between layers, which means that the potential which pins the Na charges does not involve very deep minima. Our analysis of the EFG on the Co sites also reveals that the lattice contribution to the EFG is quite smaller for this phase than for the phases with lower Na content.

We therefore are lead to suggest that all these phenomena are related and that the lower structural stability and the difference in the electronic properties are linked. The magnetic order being robustly found at $T_{N}=22$~K by most authors, for samples of different qualities, appears insensitive to the actual stacking of the Na planes. This would mean that the overall AF interaction between Co layers is rather well defined and not dependent on the actual Na 3D order. On the contrary, this would surprisingly mean that the perfect 3D order of the Na found in the $x=2/3$ phase would lead to a weak magnetic interaction between Co layers. This would be required to explain the absence of magnetic order and the large variation of spin susceptibility found down to $T=50$~mK in
that phase.\cite{EPL2008}

As for the electronic properties, we evidenced that the hole contribution to the EFG and NMR shifts on the Co sites are much smaller than expected for a single $t_{2g}$ orbital. The magnitude of the contribution per hole on the Co does not appear to change significantly with Na content. This implies that the hybridization of $t_{2g}$ orbitals involved in the electronic structure does not change markedly with hole doping of the Co band. This would indicate that although the Na ordering probably permits to pin the charge disproportionation and to reveal it through our NMR/NQR techniques, it does not play a large role in the electronic structure, which might be an intrinsic property of the CoO$_{2}$ layers. We believe that this result simplifies the modelling of the electronic structure and that computations involving both the LDA starting point and taking into account the electronic correlations based on DMFT would be able to better describe such an experimental situation.

\section{Acknowledgments}

We would like to thank here F.~Bert, P.~Mendels and J.~Bobroff for their
help on the experimental NMR techniques and for constant interest and
stimulating discussions. S.A.K is indebted to L.R.~Tagirov for elucidating
communitations. I.R.M. thanks for the support of a visit to Orsay by
``Investissements d'Avenir'' LabEx PALM (ANR-10-LABX-0039-PALM). This study
was partially supported by the Russian Foundation for Basic Research
(project 14-02-01213a) and by the Program of Competitiveness Growth of Kazan Federal University funded by the Russian Government. The work of S.A.K. was supported by the Ministry of Education and Science of the Russian Federation (state order 2014/57).

\appendix

\section{NMR spectra for $H\parallel c$ and $H\perp c$}

\label{AppCoSpectra}

\begin{figure}[tbp]
\includegraphics[width=1.0\linewidth]{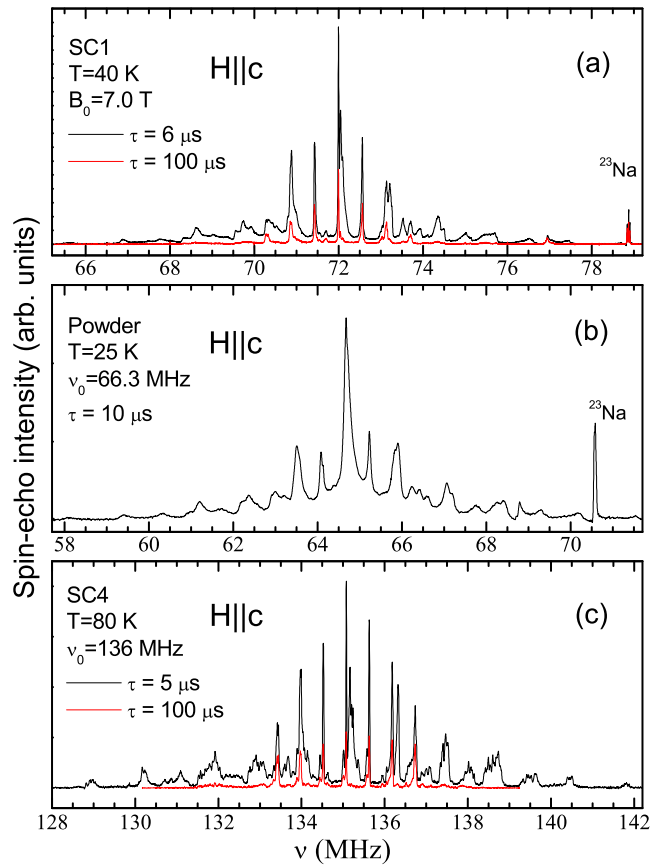}
\caption{(Color online) $^{59}$Co NMR spectra taken with the applied
magnetic field $H\parallel c$ in different samples and experimental
conditions: (a) Single crystal SC1 measured at T=40~K in the fixed magnetic
field $B_{0}=$7~T. Spectra measured at short time interval $\protect\tau $ =
6$\protect\mu $s between rf pulses (black line) and long $\protect\tau $ =
100$\protect\mu $s (red line) are shown; (b) Powder sample measured at
T=30~K, $\protect\tau $ = 10$\protect\mu $s and in sweep field mode with $%
\protect\nu _{0}$ = 66.3~MHz; (c) Single crystal SC4 measured at T=80~K in
the sweep field mode with $\protect\nu _{0}$ = 136~MHz. Spectra measured at $%
\protect\tau $ = 5$\protect\mu $s (black line) and $\protect\tau $ = 100$%
\protect\mu $s (red line) are shown.}
\label{FigCo77Co40KHc}
\end{figure}

Figure~\ref{FigCo77Co40KHc} shows the NMR spectra measured in the two
different single crystals and in the oriented powder sample at different
temperatures with the direction of the applied magnetic field $H$ parallel
to the crystallographic $c$-axis of the sample ($H\parallel c$). The
quadrupolar satellite lines are easily seen in such spectra which reveal a
large multiplicity of Co sites with distinct EFG values. The $H\parallel c$
spectra of the single crystals samples SC1 and SC4 display identical
features as that of the powder sample, with a slightly better resolution and
slightly narrower satellite lines - see Fig.~\ref{FigCo77Co40KHc}. This
indicates that the $c$ axis orientation is better defined in the single
crystals, which were easily cleaved, while perfect alignment of the $c$ axes
of the single crystal grains of the powder is harder to achieve. Therefore
the single crystals spectra are more useful and discussed in details in this
paper.

\begin{figure}[tbp]
\includegraphics[width=1.0\linewidth]{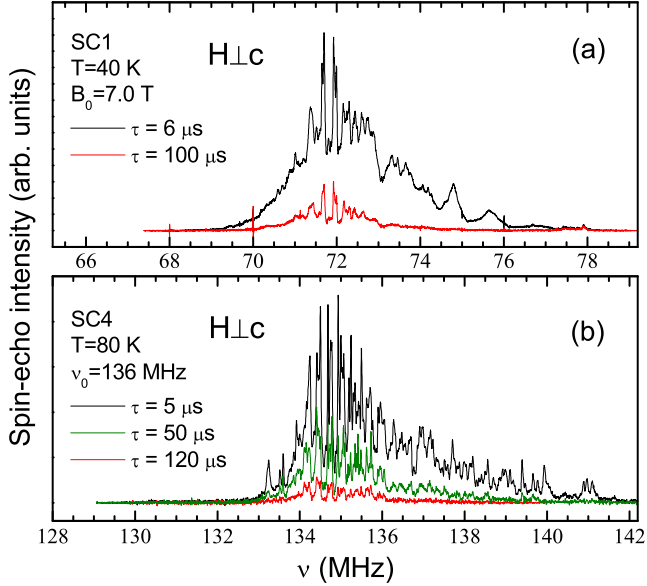}
\caption{(Color online) $^{59}$Co NMR spectra measured in $H \perp c$
orientation: (a) Single crystal SC1 measured at T=40~K in the fixed magnetic
field $B_{0} =$7~T. Spectra measured at short time interval $\protect\tau $
= 6$\protect\mu $s between rf pulses (black line) and long $\protect\tau $ =
100$\protect\mu $s (red line) are shown; (b) Single crystal SC4 measured at
T=80~K in the sweep field mode and $\protect\nu_{0}$ = 136~MHz. Spectra
measured at $\protect\tau $ = 5$\protect\mu $s (black line), $\protect\tau $
= 50$\protect\mu $s (green line) and $\protect\tau $ = 120$\protect\mu $s
(red line) are shown.}
\label{FigCo77Co40KHab}
\end{figure}

When the external field is applied in the $H\perp c$ direction the $^{59}$Co
NMR spectra are very complex, even in the single crystal samples - see Fig.~%
\ref{FigCo77Co40KHab}. As was shown from the $^{23}$Na NMR study done in
Ref.~\onlinecite{Na077prb}, the SC1 sample is not a single crystal in the $%
a-b$ plane and contains at least many twin boundaries and more probably a
mosaic of crystallites with different orientations in plane. As a
consequence in the $H\perp c$ direction the SC1 $^{23}$Na and $^{59}$Co NMR
spectra appear quite analogous to the powder sample spectra. The SC4 $H\perp
c$ spectrum has sharper features which proves more perfect in-plane order in
this sample than in SC1, but mosaicity still remains in that sample.

\section{Intensity analysis}

\label{AppIntenAnalysis}

To obtain the final simulated spectrum, which is shown by red line in Fig.~%
\ref{FigCo77HFSpec}, the separate contributions were summed with the weights
shown in the Table~\ref{tab:Co77Params}. These numbers were obtained by
careful studies of specific parts of the experimental spectrum. For example,
as one can see in Fig.~\ref{FigCo77HFSpec} the two well resolved lines in
the low or high frequency parts of the spectrum corresponds to the $\pm 
\frac{5}{2}\leftrightarrow \pm \frac{7}{2}$ satellite transitions of the
Co3a ($\approx $130.9~MHz and $\approx $141.3~MHz) and Co3b ($\approx $%
129.6~MHz and $\approx $142.8~MHz) NMR signals. To obtain their relative
intensities they were integrated and the results were corrected by the
transverse nuclear magnetization relaxation effects ($T_{2}$ correction).
This analysis gives an intensity ratio Co3a/Co3b=2/1 with better than 5\%
accuracy. The $-\frac{3}{2}\leftrightarrow -\frac{5}{2}$ transitions of the
Co2c ($\approx $138.8~MHz) group are also well separated in the $^{59} $Co
NMR spectrum (see Fig.~\ref{FigCo77HFSpec}). Taking into account the
relative intensities of the $\pm \frac{3}{2}\leftrightarrow \pm \frac{5}{2}$
and $\pm \frac{5}{2}\leftrightarrow \pm \frac{7}{2}$ quadrupolar satellites
and the $T_{2}$ corrections we deduced that the intensities of the Co2c and
Co3a are almost equal (with about 20~\% accuracy). Next we determined the
intensities of the $\pm \frac{5}{2}\leftrightarrow \pm \frac{7}{2}$
transitions of the Co1 (135.5~MHz and 137~MHz) which are well separated in
the spectrum measured at long time interval $\tau $ = 100~$\mu $ s between
rf pulses (see Fig.~\ref{FigCo77Co1Spectra}), and compared their intensity
with the same transitions of Co3a. After $T_{2}$ correction, this gave us
the intensity ratio Co1/Co3a=3/2.

Finally, we established earlier on the same $x=0.77$ samples that the
two-dimensional structure of the Na order corresponds to 10 Na sites on top
of a 13 Co sites unit cell.\cite{Na077prb} The intensity analysis of the
central line $(-\frac{1}{2}\leftrightarrow +\frac{1}{2})$ of the $^{59}$Co
NMR spectrum allowed us to deduce that the ratio of the Co1 to the total
cobalt NMR intensity is equal to 20(3)\%.\cite{Co77JETP} This number
corresponds very well to 3 Co sites over 13 expected for the unit cell.
Using the relative intensities which we found above we deduce that Co2a and
Co2b would correspond to the five remaining Co sites. By comparing
quadrupolar satellites for these lines we came to the conclusion that their
intensity corresponds to Co2a/Co2b=4/1, this ratio remaining the most
inaccurate in this intensity analysis and the ratio Co2a/Co2b=3/2 couldn't
be completely excluded.

\section{Quadrupolar satellites substructure}

\label{AppQuadrSatel}

As one can see in Fig.~\ref{FigCo77HFSpec} the agreement between the
simulated spectrum and the experimental one can be considered as very good.
Does that mean indeed that the structure of this $T_{N}=22$~K phase does
correspond to a single crystal unit cell including two CoO$_{2}$ planes and
13 Co per planar unit cell? The following detailed examination of our NMR
spectra and comparisons with the data on the $x=2/3$ phase lead us to
question that possibility.

\begin{figure}[tbp]
\includegraphics[width=1.0\linewidth]{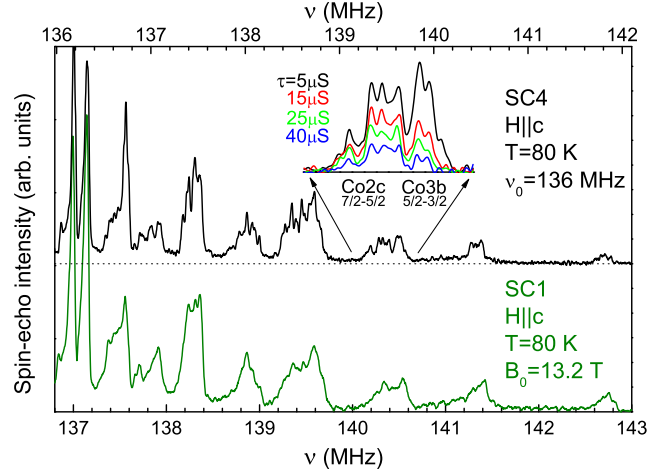}
\caption{(Color online) Comparison of the part of $H \parallel c$ $^{59}$Co
NMR spectra of the SC1 (lower panel) and SC4 (upper panel) samples measured
at $T$=80~K. In the inset of upper panel we show the same group of satellite
lines measured at different time interval $\protect\tau $ between rf pulses
which proves that the observed substructure is not associated with
experimental noise but come from effects due to differences in the stacking
of the Na layers.}
\label{FigCo77SC1SC3}
\end{figure}

In Figure~\ref{FigCo77SC1SC3} we display a magnified part of the
experimental $^{59}$Co NMR spectra in the SC1 and SC4 samples. As one can
see the quadrupolar satellites which we considered so far in Fig.~\ref%
{FigCo77HFSpec} as single lines with some noise, do in reality represent
groups of lines with slightly distinct $\nu _{Q}$ values. This is valid for
all Co2 and Co3 NMR signals considered in those spectra. Such a
sub-splitting has not been seen for Co1.

Therefore the values of the quadrupolar frequency $\nu _{Q}$ listed in the
Table~\ref{tab:Co77Params} for the various Co2 and Co3 lines should be
considered as average $\nu _{Q}$ values for groups of Co sites with quite
similar local structure and $\Delta \nu _{Q}$ represents the distribution of
the EFG values inside each group. Also the magnetic shift $K_{ZZ}$ values in
the Table~\ref{tab:Co77Params} are the values of center of symmetry position
for the quadrupolar satellites for the group of Co sites. This scattering
means that the 3D order is not perfect and that some defects are present but
that the charge disproportionation is not so much affected by the 3D order.

\section{Spin lattice relaxation of the cobalt sites}

\label{AppT1details}

\begin{figure}[tbp]
\includegraphics[width=1.0\linewidth]{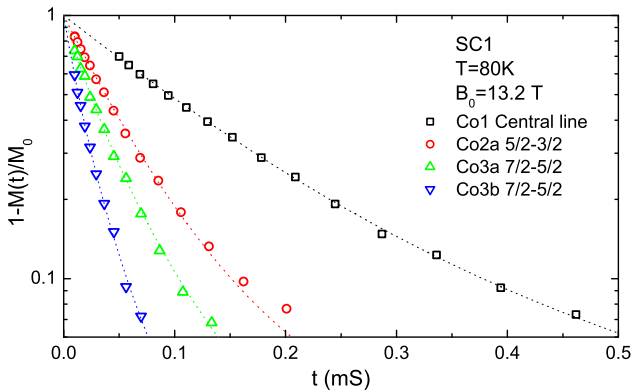}
\caption{(Color online) Spin-lattice relaxation curves measured on different
cobalt sites in SC1 sample in the same experimental conditions as spectrum
in Fig.~\protect\ref{FigCo77HFSpec}. The fits of the data with the
magnetization relaxation functions from Ref.~\onlinecite{H67_CoNMR} with
parameters from the Table~\protect\ref{tab:Co77Params} are also shown.}
\label{FigCo77T1curves}
\end{figure}

We have already shown that the magnetic sites can be differentiated by their
magnetic shifts $K_{zz}$ and by their $T_{2}$ values given in Ref.~%
\onlinecite{Co77JETP} from central line studies. Alternatively as we have
shown in Refs.~\onlinecite{H67NQRprb,H67_CoNMR} the magnetic sites can be
differentiated by their $T_{1}$ relaxation times. Due to the close values of
the magnetic shifts for all cobalts sites in the $H\parallel c$ direction
(see Table~\ref{tab:Co77Params}), to avoid cross-relaxation between sites we
took $T_{1}$ data on quadrupolar satellite transitions of different sites
which are quite separated in frequency - see for example Fig.~\ref%
{FigCo77HFSpec}. The experimental technique used for spin-lattice relaxation
measurements and the relaxation functions for the different transitions for
a nuclear spin $I=7/2$ have been reported in detail previously.\cite%
{H67_CoNMR} In Fig.~\ref{FigCo77T1curves} we show some of the longitudinal
nuclear spin magnetisation decays. There the fits of the data which permit
to determine the $T_{1}$ values are also reported. The deduced $T_{1}$
values are listed for all cobalt sites in Table~\ref{tab:Co77Params}.

Here again the Co1 has the longest relaxation time $T_{1}$ which confirms
the non-magnetic character of this cobalt site. Co3a and Co3b have rather
short $T_{1}$ values as expected for magnetic Co site behaviour, and all
Co2's have quite similar $T_{1}$ values in between those of Co1 and Co3b.

\section{O71 Phase}

\label{AppO71}

\begin{figure}[tbp]
\includegraphics[width=1.0\linewidth]{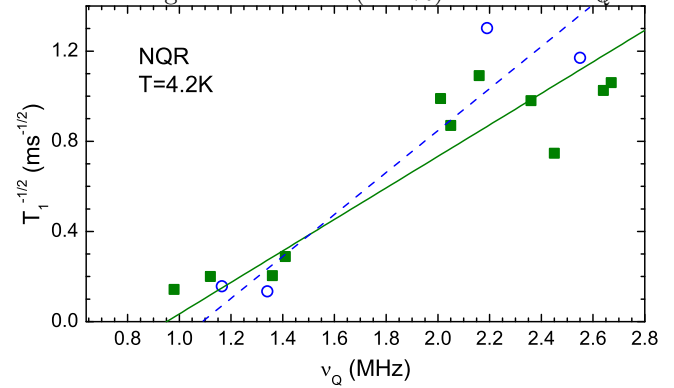}
\caption{(Color online) $(1/T_{1})^{1/2}$ vs. $\protect\nu _{Q}$ for
different cobalt sites in $x=2/3$ (open blue circles) and $x=0.71$ (green
full squares) phases measured in NQR experiments at T=4.2~K. Linear fits are
shown by solid lines with corresponding colour.}
\label{FigCo71}
\end{figure}

We have evidenced the existence of three stable phases for $x=2/3$, $x=0.71$
and $x=0.72$.\cite{EPL2008,JETP4phasesNQR} We did not study in detail the $%
^{59}$Co NMR spectrum for the two latter ones, but their difference has been
well established by NQR experiments.\cite{O71Platova} For $x=0.71$ the $%
^{59} $Co NQR data allowed us to resolve at $T=5~$K eleven Co sites with
distinct $\nu _{Q}$ values, which are different from those of the $x=2/3$
and $x=0.72$ phase.\cite{JETP4phasesNQR} So we could compare $1/T_{1}$ data
measured at 4.2~K in $x=0.71$ phase with those of the four NQR lines in the $%
x=2/3$ phase taken in similar conditions - see Fig.~\ref{FigCo71}.

We see that in the $x=0.71$ phase we have fast relaxing sites with a
significant scatter $(\pm 20\%)$ of values of $\nu _{Q}$ and of $%
(1/T_{1})^{1/2}$ but the trend of this quantity versus $\nu _{Q}$ is
identical to that found for the $x=2/3$ phase (see Fig.~\ref{FigCo71}).
Furthermore the numerical values are very similar for the two phases. Here
again this means that the $\nu _{Q}$ values distinguish 11 sites, but only
two markedly distinct magnetic properties similar to Co1 and Co2 in the $%
x=2/3$ phase are revealed from the $(1/T_{1})^{1/2}$ versus $\nu _{Q}$
plots. If we just restrict the data to the weighted averages of these Co1
like and Co2 like sites one gets from the NQR data a vanishing value of $%
(1/T_{1})^{1/2}$ for $\nu _{0}\simeq 0.95$~MHz which would be associated
with the non magnetic Co$^{3+}$. In the same experimental conditions the NQR
data for $x=2/3$ yields as well $\nu _{0}\simeq 1.08$~MHz in agreement with
the value obtained from the higher temperature NMR data. A correction for
the orbital contribution to $T_{1}$ is not required in that case as the spin
term contribution increases at low $T$ and becomes dominant.

Here the slope of linear fits is larger by a factor 1.4 for $x=2/3$ with
respect to $x=0.71$. This is consistent with the ratio of 1.6 found at 5~K
from $^{23}$Na NMR data for $(1/T_{1})^{1/2}$.\cite{EPL2008}

% Produces the bibliography via BibTeX.
\bibliography{NaxCoO2}

\end{document}